\documentclass{bmcart}[linenumbers]

\usepackage{hyperref}
\usepackage{graphicx}
\usepackage{multicol}

\usepackage[utf8]{inputenc} 

\startlocaldefs
\endlocaldefs

\newcommand{\revision}[1]{\textcolor{black}{#1}} 

\begin{document}

\begin{frontmatter}

\begin{fmbox}
\dochead{Research}

\title{Deflating the Chinese Balloon: Types of Twitter Bots in US-China balloon incident}


\author[
   addressref={aff1},                   
   corref={aff1},                       
   email={lynnetteng@cmu.edu}   
]{\inits{LHXN}\fnm{Lynnette Hui Xian} \snm{Ng}}
\author[
   addressref={aff1},
   email={kathleen.carley@cs.cmu.edu}
]{\inits{KMC}\fnm{Kathleen M} \snm{Carley}}


\address[id=aff1]{
  \orgname{IDeaS Center for Informed Democracy \& Social-CyberSecurity, Carnegie Mellon University}, 
  \city{Pittsburgh, Pennsylvania},                       
  \cny{United States}                                    
}

\end{fmbox}


\begin{abstractbox}

\begin{abstract}
As digitalization increases, countries employ digital diplomacy, harnessing digital resources to project their desired image. Digital diplomacy also encompasses the interactivity of digital platforms, providing a trove of public opinion that diplomatic agents can collect. 
Social media bots actively participate in political events through influencing political communication and purporting coordinated narratives to influence human behavior. This article provides a methodology towards identifying three types of bots: General Bots, News Bots and Bridging Bots, then further identify these classes of bots on Twitter during a diplomatic incident involving the United States and China. In the balloon incident that occurred in early 2023, where a balloon believed to have originated from China is spotted across the US airspace. Both countries have differing opinions on the function and eventual handling of the balloon. Using a series of computational methods, this article examines the impact of bots on the topics disseminated, the influence and the use of information maneuvers of bots within the social communication network. Among others, our results observe that all three types of bots are present across the two countries; bots geotagged to the US are generally concerned with the balloon location while those geotagged to China discussed topics related to escalating tensions; and perform different extent of positive narrative and network information maneuvers. \revision{The broader implications of our work towards policy making is the systematic identification of the type of bot users and their properties across country lines, enabling the evaluation of how automated agents are being deployed to disseminate narratives and the nature of narratives propagated, and therefore reflects the image that the country is being projected as on social media; as well as the perception of political issues by social media users.}
\end{abstract}



\end{abstractbox}

\end{frontmatter}



\section{Introduction}
Digital diplomacy can be understood as the use of digital resources by a country to achieve its foreign policy goals and proactively manage its image and reputation \cite{adesina2017foreign}. As the world becomes more digital, there is a growing us of social media platforms by countries as tools of communication with the general public, where nations tailor foreign-policy and nation-branding messages to the digital audience \cite{manor2015america}. Digitalized public diplomacy also includes the emphasis on the interactivity of digital platforms \cite{huang2020facebook}. Online political participation by individual users can provide an indication on the perception of political discussions. As individual users share their opinions on global affairs, diplomatic agents can systematically collect these information for campaign analysis \cite{theocharis2018continuous,ng2023combined}.

Twitter has become a popular platform for political discussions. Its fluid microblogging structure promotes diversity of opinions and provides a digital platform for messaging campaigns to cross geographical boundaries \cite{benney2011twitter}. Within this digital space, social media bots have been observed to be involved in cross-country relations within the digital space. Bots are \revision{loosely defined as inauthentic social media users (i.e., fake user profiles) \cite{boshmaf2011socialbot}, who are sometimes} controlled through software automation to post content or interact with other users \cite{gorwa2020unpacking}.

Bots are under scrutiny because past studies have observed that they can be used to alter perceptions of political discourse on social media. Bots associated with Russia have been observed to have employed tactics to sow discord and support specific candidates during the 2016 US Presidential elections \cite{linvill2019russians}, which may have contributed towards the perception of American citizens towards the candidates.  This is especially of concern if the the bots participate in diplomacy, which is the art of ``influencing the decisions and behavior of foreign governments and peoples through dialogue", according to Encyclopedia Britannica \cite{britannicaDiplomacyDefinition}. 
In the 2022 Russia-Ukraine war, bots were deployed by both countries on Twitter to shape support for the war. Ukrainian bots overwhelmed the conversation in tweet quantity, but Russian bots had more effective communication manufacturing conflict \cite{shen2023examining,zhao2023manufacturing}. Russia bots took part in extensive agenda building activities during the 2016 US elections, showcasing how these automated accounts can be used not only to brand messages for domestic audiences but also for foreign audiences \cite{linvill2019russians}. \revision{While these accounts are inauthentic, not all these accounts were fully controlled by software: the Russian Internet Research Agency conducted their work with a mixture of bots and real people employed in a St. Petersburg ``troll factory" operated by a large group of human operators \cite{dawson2019russia}. China, similarly, maintains many inauthentic accounts through the 50-Cent Party, the Chinese government's campaign to shape global narratives \cite{diresta2020telling}. The commonality of these accounts in general is that they are inauthentic users and generate a large volume of mostly political tweets, flooding the social media information system.  }

The United States (US) and China are two major powers on the world stage and have been locked in periods of high tensions. One observation of digital diplomacy between the two countries occurred in 2021, where the US released information on the tracing of the origins of COVID-19, Chinese diplomatic Twitter accounts asserted support for scientific tracing and opposed the politicization of the tracing \cite{repnikova2023asymmetrical}. Anti-Chinese state political views were discovered on microblogging platforms Twitter and Weibo over the 2017 Spring Festival period. These narratives also include pro-Hong Kong and pro-Uyghurs independence themes, suggesting algorithmic manipulation to boost democratic ideals \cite{bolsover2019chinese}.

One cross-country diplomatic incident between the US and China arose in early 2023. In January 2023, a balloon was spotted floating around in American airspace. US officials quickly determined that it was a high-altitude surveillance balloon originating from China. The balloon was estimated to be flying as high as 60,000 feet above ground, which puts it at about ten times closer than the lowest Earth-orbiting satellites \cite{barnes2023china}. China asserted that the balloon was merely a weather balloon. The US tried to open lines of communication with Chinese President Xi Jinping, and China expressed its dissatisfaction with the accusation of surveillance \cite{UrgedReflect}. Both countries maintained strong stances with regards to the functionality of the balloon. The balloon was eventually shot down by a US F-22 Raptor fighter jet on February 4, 2023. Online, this balloon incident drew much attention which inspired discussions and memes of fear, anger and humor \cite{kutllovci2023s}.

This study seeks to study the online political participation during the balloon diplomatic incident between US and China. It measures the inauthentic online engagement aspect of digital diplomacy through analyzing the scope of bot activity on social media. It does so by examining three different types of bots: \texttt{General Bots} that are identified by generic bot detection algorithms, \texttt{News Bots} that play a key role in providing automated updates to news stories; and \texttt{Bridging Bots} connect communities with each other. \revision{Briefly, users that are identified as bots through a bot detection algorithm are extracted out. \texttt{News Bots} are identified through containing the word ``news" in their profile, or having 90\% of their tweets classified as news headlines through a machine learning classifier. \texttt{Bridging Bots} are identified through their position in a all-communication social network, where they are the bots that straddle between two algorithmically determined Louvain clusters. \texttt{General Bots} are bots that are not identified as any of the other two bot types. }

This analysis of bots provides insights to inauthentic and automated activity during a diplomatic event that may influence political perceptions of human users. While bots used in digital campaigns are not a new phenomenon, the examination of the social roles that bots play online within a diplomatic incident have not been fully studied. \revision{In this work we examine the social roles of different types of bots per region in the Chinese balloon diplomatic incident to provide insights towards the country's portrayal and social media sentiment in the US-China relationship.} 

\subsection{Contributions}
Through combining social network analysis, topical analysis and sentiment analysis, we form a picture of the position and perspectives of these automated accounts. Specifically, the contributions of our paper are three-fold:
\begin{enumerate}
    \item We study a subset of digitalized public diplomacy through the analysis of inauthentic online engagement during a diplomatic incident involving two major world powers.
    \item We define and develop methodologies to identify three key types of Twitter bots: General Bots, News Bots and Bridging Bots. 
    \item With these techniques, we analyze the presence of the three types of bots within a discussion of an event that involved two countries, the Balloon incident between US and China. We do so by analyzing the topic, network and information maneuver techniques of the different types of bots, separated by their self-declared geographic locations.
\end{enumerate}

From this investigation of digital diplomacy on Twitter, our results show that the three types of bots (General Bots, News Bots and Bridging Bots) are consistently present throughout the three geographies (US, China, Rest of the World). The communication network formed by users depict that users from the US and China naturally form separate clusters. Bots geotagged to US are generally concerned with the location of the balloon and the possible surveillance functionality, while bots geotagged to China discussed topics on war and escalation. \texttt{General Bots} and \texttt{News Bots} are dispersed throughout the network, while \texttt{Bridging Bots} are rarer and typically connect between users of both countries. \revision{\texttt{General Bots} and \texttt{News Bots} geotagged to China have the highest average eigenvector and centrality values, indicating that they are more well-connected within the social network compared to the US-geotagged bots, thus having a higher amount of influence. In terms of information maneuvers through the BEND framework, we observe that \texttt{General Bots} and \texttt{Bridging Bots} make use of emotional appeal rather than logical appeal, and \texttt{News Bots} often make their news headlines short and catchy to excite readership. }


\section{Related Work}

\subsection{Digital Diplomacy on Social Media}
The US and China have had a prolonged period of diplomatic tensions in both offline and online world. Since 1949, both countries have had periods of cooperation, competition and tensions in issues related to military, technology, trade, climate change and views on governance (e.g., independence of Taiwan, activism within China) \cite{cfrTimelineUSChina}. 

In today's information age, international politics also hinges on how each country presents the dispute in both traditional and digital media, and how the public reacts to their stories. Both the United States and China have used Twitter as a medium to present their agendas during the South China Sea dispute \cite{guo2019whose}. Strategic narratives were also dispersed on Twitter by the Chinese authorities during the 2020 COVID-19 pandemic to frame the government response to the pandemic as a proof of the country's strength and resilience \cite{moral2023assembling}. Public responses to politics can also be measured through Twitter discourse, for example the examination of emotional tendency of online users towards the presence of US warships in the South China Sea \cite{xiaomeng2020emotional}.

The balloon incident comes at a critical moment in US-China relations. Representatives from both countries have met in late 2022 to agree to deepen bilateral relations. They had planned to meet again in January 2023, but the meeting was canceled due to the balloon incident \cite{kutllovci2023s}.
The increased tensions during the balloon incident is not due to the surveillance functionality of the balloon, but rather, is emblematic of the rising tensions between the two countries over time \cite{kutllovci2023s}.  The presence of surveillance by China indicates her desire to collect information on its geopolitical rival amid the frayed diplomatic relations. 

Chinese government messaging has always leaned towards nation-building and regime legitimacy. The Chinese Communist Party (CCP) has employed the use of emotions such as fear, rage and pride to exert an influence on contemporary Chinese politics \cite{zhang2023despicable}. Studies have observed that the younger Chinese generation embody a nationalistic sentiment as a result of \revision{being surrounded by Chinese nationalism emotions seeded in official Chinese propaganda online\cite{guo2004cultural}.}

\revision{For the United States, information cyber operations encompasses, among others, the art of spreading propaganda and information to change human behavior \cite{aleroud2023span}. Past work have observed that routine US government videos generally revolve around social issues such as of law enforcement, social issues, transportation, economic development and political issues \cite{10.1145/2037556.2037574}. Recruitment-intended publicity videos identify with professional and fitness careers such as students, doctors and athletes \cite{10.1145/3617127}. While Chinese government messaging aims at enhancing the viewer's love for the country, US government messaging aims to address social issues. }

\revision{In terms of analyzing differences of the discussions users engage in through patterns of online firestorms between the United States Twitter and China Weibo, past work observed that firestorms in both platforms have significant cultural differences. Users from the two regions US and China engage on social media platforms with different communication behaviors. China Weibo tend to target the social responsibility or ethics-related dimensions, while those in US Twitter target reputation and ability \cite{kim2021online}. }

\subsection{Social Media Bots}
\label{subsec:lit_bots}
Social media bots\revision{, which are inauthentic or automated-like accounts, have been observed to participate in political discussions}. Evidence of social media bots attempting to manipulate political communication dates back to as early as 2010: in the 2010 US midterms elections, bots were employed to support and discredit candidates by injecting thousands of Tweets pointing to websites with fake news \cite{bessi2016social}. Similarly, in the 2016 US Presidential campaign, there are two faction of bots supporting both presidential candidates, which run both supportive campaigns for the candidate they affiliate with, and smear campaigns for the opposition candidate. These bots are centrally embedded within the social network, giving them a huge amount of influence \cite{bessi2016social}. During the same elections, bots originating from Russia actively sowed discord in a complex multi-platform disinformation campaign, purchasing advertisements to promote their political ideals and disseminating curated news content \cite{ng2021does}.

Bots that seek to purport coordinated narratives to influence human behavior are more prevalent in politics rather than other realms like art and sports \cite{stieglitz2017social}. Groups of bots working together have been observed to change the stances of human users online \cite{ng2022pro,gionis2013opinion}. This is an area of note because changes in a person's stance can be detrimental to society if the person has been convinced to do public harm. These have led to increase coverage and research on the impact of social media bots on the political scene.

Social media bots have been observed to perform information maneuvers. Governments like those of Iran and China are using social media automation to disseminate propaganda and provide digital entertainment to their population \cite{aleroud2022theory}. Russian bots have been identified to pretend to be English speakers and exhibit hostile attitudes towards political opponents and Western democracies, using persuasive information maneuvers to express skepticism and promote a lack of trust in the existing governments \cite{alieva2022disinformation}.

To identify whether an account is a bot or human, a series of bot detection algorithms have been developed in literature. These algorithms evaluates the features of social media accounts (e.g., user name, account metrics) and returns a likelihood between 0 and 1 representing whether the account is likely to be a bot or human \cite{ng2022stabilizing}. Bot detection algorithms are generally machine learning algorithms that train on manually annotated datasets that indicate bot/human labels for data points. These algorithms range from random forests classifiers \cite{schnebly2019random} to neural network formulations \cite{ng2023botbuster} to deep learning architectures \cite{martin2021deep}. 

The training datasets of these bot detection algorithms consists of a wide variety of bot types from different countries. These algorithms are constructed as generic algorithms that can be trained and applied across multiple datasets with reasonable accuracy. For example, a random forest classification model has had an average performance of 0.72\% accuracy of across 11 datasets \cite{yang2020scalable}, and neural network-based models constructed on a large proportion of Twitter datasets can also be applied on Reddit datasets with 69.77\% accuracy \cite{ng2023botbuster}. Therefore, in this paper, we make use of a pre-trained Twitter bot detection algorithm, BotHunter \cite{beskow_bot}, that should be generalizable across the training dataset and the political discourse of our dataset. 


\section{Data}
\subsection{Data Collection from Twitter}
We tracked the political discourse of the Chinese balloon on Twitter. Using a Python script, we collected tweets using the Twitter V1 Streaming API. We streamed tweets containing the search terms \#chineseballoon and \#weatherballoon from 31 Jan 2023 to 22 Feb 2023. \revision{From our streamed tweets, we filtered for tweets only in the English language.} In total, we collected 1,192,445 tweets from 121,048 unique users.

\subsection{Geographic location identification}
We aim to separate the accounts into the two major powers involved in this incident: United States (US) and China, by using information disclosed by the user account. We do this through geographic location identification techniques.

Mapping the geographic location of Twitter users have been a studied topic. Since late 2009, Twitter allowed users to include geographic location as metadata to their profile. This metadata can be included in terms of latitude/longitude coordinates or by place name (georeferenced tweets). The problem with georeferenced location, which are location expressed in text form like ``New York City", is that it is prone to duplicates and ambiguity. Geocoding algorithms thus identify location from the surrounding texts (e.g. ``Chinatown, New York City" identifies a specific Chinatown through the city name), disambiguates it and converts it to its approximate map coordinates \cite{leetaru2013mapping}. The disambiguation step involves methods such as text mapping to the Wikipedia gazetteer or a global city gazetteer \cite{leetaru2013mapping}; or estimation of a user's location based on the content of their tweets, a method that assumes users in similar regions will tweet similar trending topics \cite{cheng2010you}.

For each user account, we assign a geolocation based on their disclosed location. To do so, we extract the ``country" field from the account's meta-data and perform a reverse geolocation search using Nominatim API 4.2.1\footnote{\url{https://nominatim.org}}. We input location information from the account's metadata (i.e., latitude/ longitude, location string) to the API call. The API returns a JSON object containing location information, such as state and city, and we extract the country term from this information output. For example, if the account declared its location as ``San Francisco, California", and the Nominatim search result returns ``United States", we indicate the account to be from the country United States. Similarly, an account with the declared location ``Beijing" with the returned Nominatim result ``China" will be geotagged as from China. For accounts with locations other than US and China, we classify them into accounts from the ``Rest of the World". Accounts where the geolocation is not disclosed or that the Nominatim API does not return a result are disregarded in our subsequent downstream analysis.

Despite the use of geographic location identification, we cannot definitively claim the users originate from these countries. Twitter, as well as other social media sites like Google, Facebook, WhatsApp and YouTube, are banned by China's ``Great Firewall". Many users access the application through circumventing the blockage with VPN services \cite{techcrunchDespiteBan}. In this study, we do not make a distinction of whether the users are definitely from China or are spoofing their location to be from China. Rather, we focus on the users that participate in the online discourse and present themselves to be from China, and also those that present themselves to be from the US. The external self-presentation of location by Twitter users provide an illusion of the activity from the region on the social media space which we use that in our analyses.

\section{Methodology}
\subsection{Overview}
In this study, we define and identify three types of bots -- \texttt{General Bots}, \texttt{News Bots} and \texttt{Bridging Bots}-- in the Twitter discussion surrounding the US-China balloon incident. We use a mixture of machine learning and network analysis methods to segregate these bot accounts from the general pool of accounts.
Following the identification of three types of bots, we analyze their activity in terms of their influence in the social network, the narrative themes they put forth and their expressed emotions. We compare these parameters towards Human users, characterizing the differences in the online opinions and discussions between Bots and Humans during this cross-country incident. 

For deeper analysis of bot activity within this balloon incident, we used three parameterizations: social network analysis, topical analysis and sentiment analysis.

\subsection{Identifying Types of Bots}
In this study, we identify three types of bots:
\begin{enumerate}
    \item \texttt{General Bots}: bot accounts that are identified using generic bot detection algorithms
    \item \texttt{News Bots}: bot accounts that spread news information, whether through posting original news or by aggregating news accounts
    \item \texttt{Bridging Bots}: bot accounts that build a communication pathway between two clusters of users
\end{enumerate}

\begin{figure}[h!]
\includegraphics[width=0.95\textwidth]{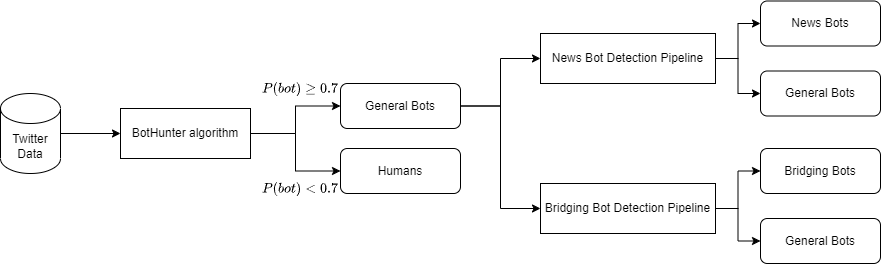}
\caption{Overview of Methodology for Identifying Types of Bots}
\label{fig:overallflowchart}
\end{figure}

In the following subsections, we describe our methodology for identifying these types of bots in detail. We note that the classification of an agent into a bot is non-exclusive. For example, a bot user can be both a \texttt{bridging bot} and a \texttt{news bot}. However, for the purposes of this study, if a user is classified as a \texttt{news} or \texttt{bridging bot}, we do not consider it to be a \texttt{general bot}. This construction allows us to perform more segregated analyses towards each bot type. The overview of the identification of the three types of bots are \autoref{fig:overallflowchart}.

\subsubsection{General Bots}
Bot detection algorithms (see \autoref{subsec:lit_bots}) are constructed to be generic models and be able to identify a broad range of Twitter bots. As such, we term the bots that are identified through these bot detection algorithms as \texttt{General Bots}. 

In this study, we identified \texttt{General Bots} using the BotHunter algorithm \cite{beskow_bot}. This algorithm uses a tier-based random forest structure to return the probability of the account being a bot. The algorithm uses an account's tweet text information, temporal patterns of tweeting, and user profile information to determine whether the account is a bot or human. The classifier performed with  $\sim 98\%$ accuracy on the original dataset it was constructed on \cite{beskow_bot}. This classifier has since been used in studies that studied bots conversations on diplomatic incidents between countries, such as China and Taiwan \cite{jacobs2022taiwan}, and Russia and Ukraine \cite{alieva2022investigating}.


We ran all user data through the BotHunter algorithm, retrieving the bot probability scores for each user. The scores were in the range of 0 and 1. \revision{After retrieving the bot likelihood score for each user, one typically sets a threshold value, above which the user is deemed as a bot, while below which it is deemed as a human. In literature, a variety of threshold values have been set, ranging from 0.25 \cite{zhang2019whose} to 0.50 \cite{varol2017online} to 0.76 \cite{rauchfleisch2020false}.} We adopted the 0.70 value to threshold the likelihood score for marking social media bots \revision{to be in consistent with some past work that have used the same algorithm \cite{ng2023combined,ng2022pro}}: a user with a score above or equal to 0.70 is deemed as a bot and an account below that value is deemed as a human \cite{ng2022stabilizing}. We denote the likelihood score as P(bot). The BotHunter algorithm also allows us to process historical data that has already been collected rather than requiring live access. In this article, we also refer to the \texttt{general bot} as a \texttt{bot}.

\subsubsection{News Bots}
\begin{figure}[h!]
\includegraphics[width=0.95\textwidth]{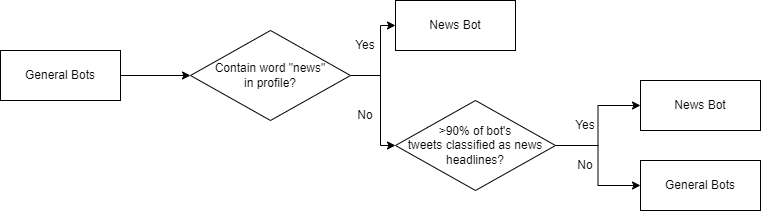}
\caption{Methodology for identification of \texttt{News Bots}}
\label{fig:newsbotflowchart}
\end{figure}

\texttt{News Bots} are bots that spread news information. These bots can either be posting original news, from which news originate from; or aggregating news from a series of other original news websites or users. We identify \texttt{News Bots} in two manners: through substring matching and via a machine learning classifier. \autoref{fig:newsbotflowchart} presents a flowchart of the methodology used to determine if an account is a \texttt{news bot}. 

The first manner relies on the explicit expression of a news bot from the user's profile information. From the set of \texttt{General Bots}, we extract the bots that contain the word ``news" in their profile (i.e., user name, screen name, description) through regex substring matching in Python. These bots are then reclassified as \texttt{News Bots}.

The second way of identifying \texttt{news bots} deals with the accounts that do not explicitly state the word ``news" in their profile but tweet news headlines. To identify these bots, we trained a random forest machine learning classifier. To do so, we obtained 100,000 examples of news headlines used between 1 January 2010 to 31 January 2020 from the News on the Web corpus \cite{now}. We enhanced the news dataset with a non-news dataset of 100,000 tweets from human users obtained from 3,298 users that includes ourselves, friends within our social network, politicians and celebrities. This classifier was implemented using the scikit-learn package in Python\footnote{\url{https://scikit-learn.org/stable/}}. This binary classifier takes in a sentence and returns whether the sentence is likely to be a news headline or not. The classifier achieved a 92.3\% accuracy. A positive classification of a tweet by this binary classifier indicates that the tweet is likely to be a news headline. By extension, if a majority of a user's tweets are news headlines, the user is likely to be a \texttt{news bot}. In this study, we use a 90\% threshold: if a classifier reflects that 90\% of a user's tweets are similar to news headlines, then we reclassify the user as a \texttt{news bot}. We use the threshold 90\%  to match the accuracy of the classifier. Then, for each bot user that has not been reclassified as a \texttt{news bot}, we run their tweets through the classifier and extract the users that are \texttt{news bots}. 

\subsubsection{Bridging Bots}
\begin{figure}[h!]
\includegraphics[width=0.95\textwidth]{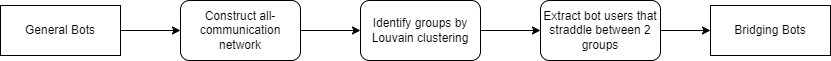}
\caption{Methodology for identification of \texttt{Bridging Bots}}
\label{fig:bridgingbotsflowchart}
\end{figure}

\texttt{Bridging Bots} build a pathway through two different groups. In this context, the two key groups are: (1) the group of users that are geolocated in the US and (2) the group that are geolocated in China. \autoref{fig:bridgingbotsflowchart} illustrates the process of identifying \texttt{Bridging Bots} in a flowchart.

We identify \texttt{Bridging Bots} in the following manner: first, we construct an all-communication social network between all the users in the dataset. In this network, the nodes represent Twitter users. Two nodes are joined together by a link if the two users have a communication relationship; that is, a retweet, @mention, quote tweet or reply tweet. Then, we perform the Louvain clustering algorithm on the constructed network to identify groups of users. From the outcome of the clustering, we identify the bot users that straddle between two clusters, and reclassify these bots from \texttt{general bots} to \texttt{bridging bots}. 

\subsection{Analyzing Twitter Bot Activity}
After classifying Twitter users as each type of bot, we analyze the activities that these bot types take part in on social media, and compare them with respect to the human users. Using a combination of computational tools, we measured the influence and emotions of the bots, as well as the themes of the narratives they are disseminating. 

In this section, we describe our methodology towards analyzing the political participation of inauthentic accounts on Twitter. The Twitter bot activity is profiled through network, topic and information maneuver analysis. \revision{We measure the difference in social network influence between bot types using network centrality measure analysis, visualize the difference and similarities of the texts put forth using topic analysis and quantitatively measure the extent of information maneuver using the BEND framework analysis \cite{BEND}.}

\subsubsection{Measuring influence through social network analysis}
Social network analysis provides a way to measure the influence of accounts through their positions in an interaction network. This analysis can provide insights to how well-connected and deeply embedded within a social network. Past studies revolving around political discourse on Twitter have observed that as the size of the communities increase, bots become more central in the rebroadcasting network, thus having a higher ability to perform influence activities \cite{bessi2016social}. Within the discourse surrounding the Russia-Ukraine war, Russian propaganda troll bots are found to actively spread pro-Russian and anti-Ukraine narratives in their Russian tweets, shaping their reasons for the war towards the Russian citizens \cite{alieva2022investigating}. The Russian bots were also found to be central to the communication network and thus have influence over the narrative.

Network analysis techniques capture the social dynamics of communities, quantifying the influence of a node through its interactions and analyzing the connection social ties of nodes \cite{tang2009social}. \revision{These techniques have been used to capture friendship relations, country trade networks and power dynamics between mafias and great families in the Florentine era \cite{wasserman1994social}.}

To begin, we construct an all-communication network, a network graph of social media users that represent communication between the users.  Each users is regarded as a node in the network graph, and a link is drawn between two nodes if they have a communication interaction, i.e., quoted, retweeted, replied-to or @mentioned each other's tweets. The weight of the links represent the number of interactions between the two users. The graph is then pruned to remove node components with less than five nodes and links that have less than ten interactions, so that we can analyze the core structure of the network. With the assumption that similar people form cohesive structures, we use the structure of this all-communication network to identify key groups and measure the influence of the groups within the networks. 

We present the results in a network graph visualization using the Gephi software \cite{bastian2009gephi}, coloring the graphs by different segregations: by country (US, China, Others); by bot class (bot, human), by bot type (news bot, not news bot, bridging bot, not bridging bot). 

From the network graphs, we calculated the average of the betweenness, eigenvector and total-degree centrality metrics between each bot class. The betweenness centrality indicates the influence the user has over information flow within the network. The eigenvector centrality indicates how well connected the user is to other highly influential users. The total-degree centrality indicates the extent of the network that can be affected by the user due to their direct connections with the user. These metrics provide quantitative insight towards the degree of influence of the users within the communication network. 

\subsubsection{Measuring themes through topic analysis}
Understanding narrative themes in groups of texts is typically done through topic analysis. Topic analysis is a pivotal way to group the large volume of information being disseminated on social media platforms into common themes. This method has been employed to distill out the degree of support and sympathy shown by other countries towards Palestine on Twitter in the 2016 Palestinian-Israeli conflict \cite{al2019multi}.
A common method for topic analysis is Latent Dirichlet Allocation, which constructs a probabilistic model of words, sentences and therefore topics. An analysis of the Twitter discourse during the 2014-2015 crisis between Russia and Ukraine where grenades were found in Kyiv, Ukraine, identified that that deleted accounts (which are most possibly bots) shared topics related to requiring accountability from Ukraine and the accusation of intimidation by Ukraine \cite{volkova2016account}. A study on the bots in the 2020 Coronavirus pandemic used topic modeling and observed that bots mainly updated news on the pandemic and promoted good hygiene habits \cite{al2020bots}.

In our topic analysis module, we first preprocess the tweet texts by removing tweet artifacts and stop words. Tweet artifacts are hashtags, URLs and @mentions that users use as part of their interaction towards others within the network. Stop words are common words like ``a", ``the", ``of", and also common event-specific phrases like ``united states" and ``china". These sets of text are removed from the tweet as they do not contribute to the overall narrative of the tweet and induce noise. 

Following which, we used sklearn CountVectorizer function\footnote{\url{https://scikit-learn.org/stable/modules/generated/sklearn.feature_extraction.text.CountVectorizer.html}} to convert the collection of tweet texts from each region into a matrix of counts that represent the frequency of tokenized words. Then, we construct word clouds to visually aid interpretation of the top 500 words. The larger the size of the word, the more frequent it is used within the text collection. Using the word cloud display, we can interpret the prevailing narratives that are expressed by each group of social media accounts. We opted to use a word cloud built on singular words as after the preprocessing step, the number of words within each tweet are quite little. Tweets are by nature short texts generated through microblogging, and have a maximum of 280 characters. The average text length after the preprocessing step is  6.3$\pm$10.6 words.

\subsubsection{Measuring information maneuvers through the BEND framework}
\revision{Information maneuvers on social media are the strategies used in the manipulation of the diffusion of information to steer mass thinking \cite{al2016understanding}.} Several frameworks that have been conceptualized to characterize information maneuvers on social media space. The BEND framework measures narrative and network maneuvers \cite{BEND}, the SCOTCH framework provides a summary of the contribution of social media actions towards the overall campaign \cite{blazek_2021}; and the ABCD framework describes in detail the Actors, Behaviors, Content and Distribution in an information maneuver \cite{alaphilippe_2020}.

In this study, we make use of the BEND framework. The BEND framework presents online campaigns as sets of narrative and network maneuvers carried out by users engaging in the social network environment, with the intent of influencing the topic-oriented communities \cite{BEND}. This framework has been applied in understanding the influence of bot users in expressing political opinions regarding the Palestine-Israel conflict \cite{danaditya2022curious}, and the Russian portrayal of one of its opposition leader \cite{alieva2022disinformation}.

We use this framework because it has a quantitative output for empirical comparison and analysis, \revision{thereby giving a repeatable output rather than a subjective classification}.
We generate probabilities of each type of bot performing the maneuvers using the ORA software\footnote{\url{http://www.casos.cs.cmu.edu/projects/ora/software.php}}. This software takes in a Twitter output file and provides empirical probabilities of the BEND cues through a weighted average of the linguistic cues derived from the text of the tweets for the narrative maneuvers, and the network cues derived from a user's surrounding all-communication network. 

Within this study, we focus on the positive maneuvers only, \revision{because diplomacy are generally image-enhancing endeavors \cite{sterling2018new}. Nations typically attempt to portray a good image towards the world \cite{surmacz2016new}, or soften their brand name (i.e., Israel's foreign ministry) \cite{manor2015america}.} Therefore, positive maneuvers are more prominent in the analysis. \revision{Positive maneuvers refer to the information maneuvers that are concerned with increasing narrative support and enlarging network groups. This does not necessarily point to good or bad implications of the text in the post, but rather it is a positive maneuver to the user in focus, expanding its reach and information dissemination patterns.} \revision{In the BEND terminology, the positive} maneuvers are the B- and E- maneuvers. The probabilities returned by the positive B- and E- maneuvers are non-negligible while those returned by the negative N- and D- maneuvers are near-zero.  In our analysis, we use the mean and standard deviation scores for each maneuver of the bot users and present the result per bot type and country.

The B- maneuvers are four positive maneuvers towards the social network: Back, Build, Boost and Bridge. These maneuvers are concerned with enhancing the connections of the user among the social community. The Back maneuver supports a narrative through increasing the number of likes, shares or @mention/reference. The Build maneuver works towards creating a group by measuring the extensiveness of co-mentioning. The Bridge maneuver introduces members of one group to another through @mentions, or shares hashtags between groups. The Boost maneuver aims to increase the group size through building links between members.

The E- maneuvers are four positive maneuvers towards the narrative: Engage, Explain, Excite and Enhance. These maneuvers are concerned with increasing the attention drawn towards a message within the social network. The Engage maneuver creates a personal affinity through the audience and a message by using lots of first-person pronouns. The Explain maneuver elaborates on a narrative through justification, using lots of function words and connectives. The Excite maneuver elicits positive emotions such as joy, excitement or pride using words affiliated with these emotions, positive emojis/emoticons and exclamation points. The Enhance maneuver expands on a narrative, through replying and quoting a tweet and adding supportive content such as text, images, URLs and so forth.

\section{Results}
\begin{figure}[h!]
\includegraphics[width=1.0\textwidth]{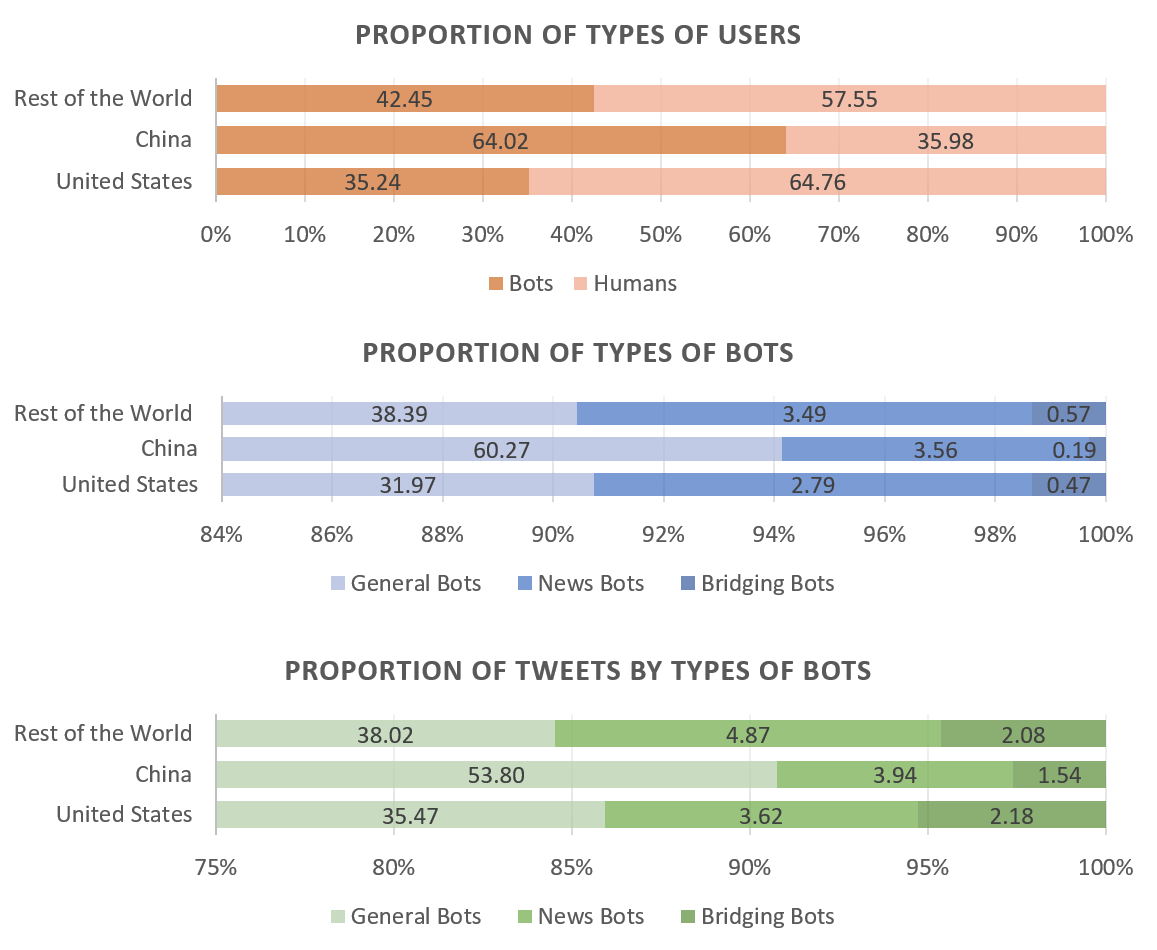}
\caption{Proportion of User Types}
\label{fig:usertypes}
\end{figure}

Within our study, we find that there are highest proportion of users geotagged to be from China that are bots, followed by the Rest of the World then the United States. These proportions are illustrated in \autoref{fig:usertypes}. In particular, for users that are geotagged as in the United States, 35.24\% (N=26,439) are identified as bots, while 64.76\% (N=48,597) are humans. For users geotagged within China, there are 64.02\% (N=5328) bots and 35.98\% (N=2995) humans. As for the proportions on the Rest of the World, 42.45\% (N=15,998) are bots while 57.55\% (N=21,691) are humans. These numbers are higher than the estimated average percentage of bots on Twitter around the world. Past studies shown that on average, 5 -14\% of the Twitter users are bot accounts, and in particular, the number of bots geotagged to the US at 14.26\% \cite{tan2023botpercent}.

The distribution of the types of bots within different regions is illustrated in \autoref{fig:usertypes}. \autoref{tab:examples} extends this illustration by providing example tweets of each bot type. \texttt{General Bots} are the most dominant types of bots from all three regions, then \texttt{News Bots}, then \texttt{Bridging Bots}. For United States, 38.39\% (N=23,992) are General Bots, followed by 2.79\% (N=2095) of \texttt{News Bots} and 0.47\% (N=352) of \texttt{Bridging Bots}. For China, 60.27\% (N=5016) are \texttt{General Bots}, followed by 3.56\% (N=296) of News Bots and 0.19\% (N=352) of \texttt{Bridging Bots}. For the Rest of the World, 31.97\% (N=14,468) are \texttt{General Bots}, followed by 3.49\% (N=1316) of \texttt{News Bots} and 0.57\% (N=214) of \texttt{Bridging Bots}. 

The proportion of tweets by types of bots is reflected in the graph in \autoref{fig:usertypes}. The number of tweets generated by each type of bot is proportional to the number of the type of bot present. For United States, \texttt{General Bots} produced 35.47\% (N=41,750) of the tweets, \texttt{News Bots} produced 3.62\% (N=4265), and \texttt{Bridging Bots} produced 0.57\% (N=2570) of the tweets. For China, \texttt{General Bots} produced 53.80\% (N=5842) tweets, \texttt{News Bots} produced 3.95\% (N=428) tweets, and \texttt{Bridging Bots} produced 1.54\% (N=2570) tweets. For the Rest of the World, \texttt{General Bots} produced 38.02\% (N=20,976) tweets, \texttt{News Bots} produced 4.87\% (N=2689) tweets and \texttt{Bridging Bots} produced 2.08\% (N=2570) tweets.

\revision{Through \autoref{fig:usertypes}, we observe that the proportion of bots is the highest in users geotagged to China, providing a glimpse into the larger amount of automation and inauthentic accounts that China-affiliated groups use as compared to US-affiliated or the rest of the world. This might reflect the extent and prolific activity of Chinese affiliated accounts targeting political issues on social media \cite{molter2020pandemics,jacobs2023tracking}. In addition, \texttt{General Bots} make up the largest proportion of bots, indicating that much of these bots do not have a specific information type to disseminate (i.e., news), or do not actively disseminate information to multiple groups (i.e., \texttt{Bridging Bots}). }

\revision{The significant proportion of bot accounts geotagged to the US indicates that the US also establishes portrayals of itself on social media accounts, using their digital diplomacy strategies \cite{manor2015america}. Digital diplomacy by the two mega-powers is crucial especially in a political event such as this balloon incident for people's perceptions of  nations are often shaped by personal encounters, especially on social media channels \cite{manor2015america}.}

\begin{table*}\centering
\begin{tabular}{|p{5cm}|p{8cm}|}
\hline
\textbf{Bot Type} & \textbf{Example Tweet} \\ \hline
\textbf{Bot} & \revision{[Rest of the World]} I'm bored why can't we have some real life alien invasion \#SpyBalloon \#ChinaSpyBalloon \#AlienInvasion \#BBCNews \#BBCNewsTen \\ \hline 
& \revision{[United States]} \#ChineseBalloon saga causing political damage already...  It's a weather balloon, so \#China said [...] (URL) \\ \hline 
& \revision{[United States]} @(User) I'm not 100\% if authentic, but it seems that is a meteorite that came down and may had produced that explosion you heard...  Unless is a fake video.  But it has nothing to do with the \#ChineseBalloon  \\ \hline 
\textbf{News Bot} & \revision{[China]} \#USA \#China \#14February [...] One could speculate that the US is using the \#ChineseSpyBalloon 'excuse' to escalate tensions with \#Beijing..  Recall that US airspace is highly controlled and that there are more accurate satellite technologies for spying (URL) \\ \hline 
& \revision{[United States]} \#USA \#China \#14February @(NewsUser)  -The \#Biden administration has blocked the sale of certain US technologies to various Chinese companies -This follows the \#ChineseSpyBalloon events  -In particular 5 companies allegedly supported programmes for these blimps [...] \\ \hline 
& \revision{[Rest of the World]} \#USA is under attack @(User). Modern wars not fought just using missiles or bombs. \#ChineseSpyBalloon \#ChinaBalloon \#Balloon could be act of Cyber war, \#Bioweapon, EMP or surveillance of nuclear installations. The world is watching how president \#Biden handle \#china firmly on this \\ \hline 
\textbf{Bridging Bot} & \revision{[United States]} @(User): Should have been shot down over Alaska. It's always about the politics. \#Biden \#ChineseSpyBalloon \\ \hline 
& \revision{[Rest of the World]} @(User1): Enjoyed joining @(User2) to discuss the \#ChineseSpyBalloon \\ \hline 
& \revision{[Rest of the World]} Balloons have been used in warfare going back 200+ years and you’re telling me the most advanced and expensive military the world has ever seen has no way to safely take out a large gently floating bag of air? @(User1) @(User2)  \\ \hline
\end{tabular}
\caption{Examples of Tweets by Bot Types. User names and URLs are removed as these can potentially uniquely identify the users. }
\label{tab:examples}
\end{table*}

\begin{figure}[h!]
\includegraphics[width=1.0\textwidth]{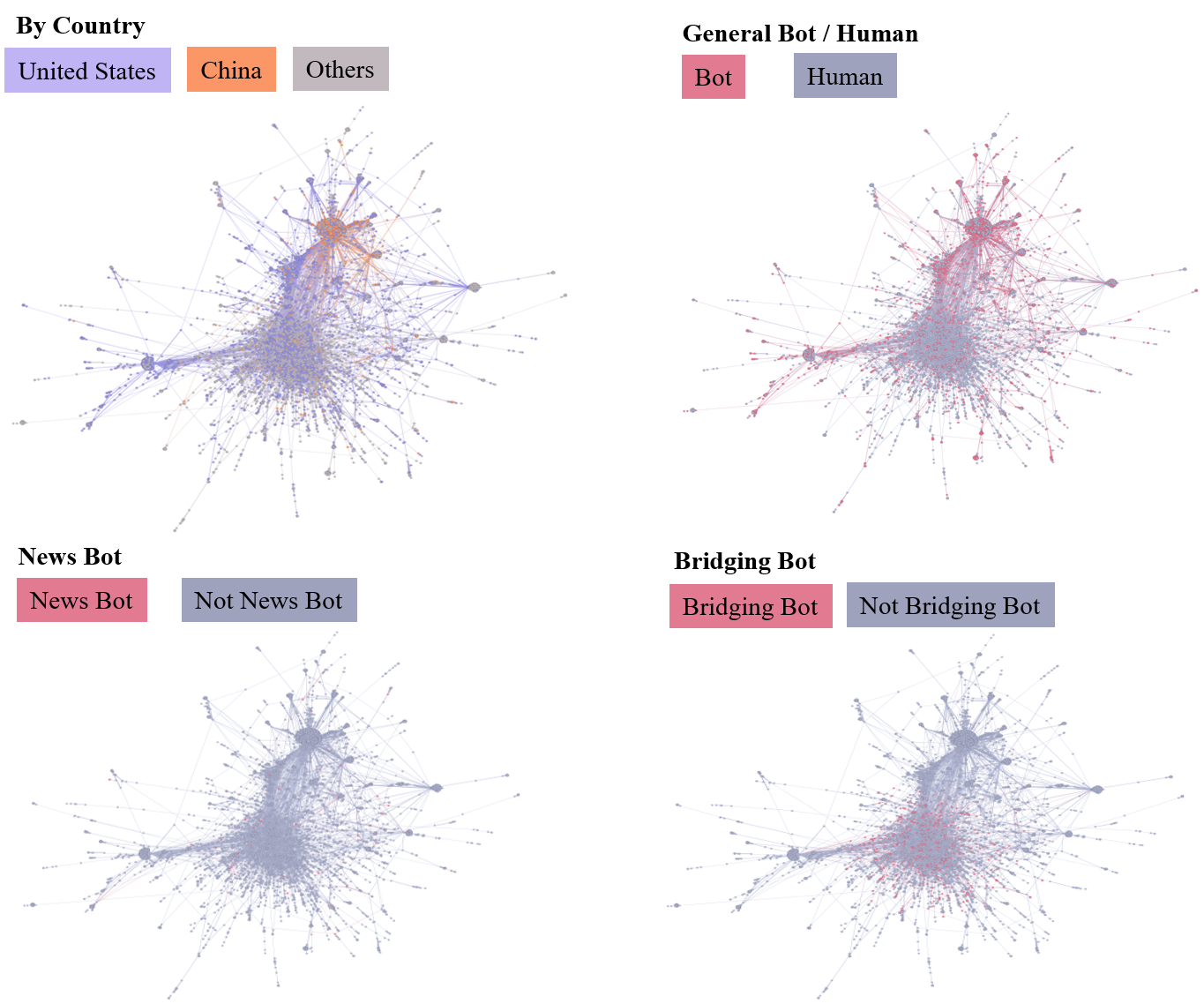}
\caption{User Type Distribution through Social Network Analysis}
\label{fig:socialnetworkanalysis}
\end{figure}

\autoref{fig:socialnetworkanalysis} depicts the distribution of user types within the all-communication social network. Users of the two key regions - US and China - are generally segregated into different clusters, indicating that they generally interact with users that are of a similar country affiliation. This demonstrates the principle of homophily, where similarity breeds connection \cite{mcpherson2001birds}. Users within each geographical cluster have an affinity to each other by means of their affiliated geolocation. Network ties of many types -- marriage, friendship, work, information transfer, and so forth -- have been observed to be homogeneous with regard to sociodemographic characteristics. Community ties that are constructed through homophily in social networks have been a source of interest, and studied for their use in network segregation \cite{bisgin2010investigating}.

\texttt{News Bots} are dispersed throughout the network, demonstrating their activity of dispensing news to a variety of user types in social media space. Given that they typically post informative news tweets, many different user types will interact with their tweets, thus they form connections throughout the network.

\texttt{Bridging Bots} are rarer throughout the network. The two distinct Louvain communities arising from this network are users from the US and users from China, respectively. Thus, \texttt{Bridging Bots} are found straddling connections between the two countries. A total of 592 accounts are identified. Such social media users are rare because they contradict the principle of homophily within the digital space, deliberately forming strong connections between multiple digital communities. \revision{The principle of homophily refers to the \revision{tendency} of similar-minded people interacting with social groups of other similar-minded people and remaining in the community \cite{khanam2023homophily}. \texttt{Bridging Bots}, however, do not entirely follow this principle since they post tweets that contain users from multiple communities, specifically @mentioning them within the post, suggesting that their choice of user tags is curated. This behavior puts them within more than one set of community, in particular communities that are disparate from each other, as observed from the Louvain clustering algorithm. } Nonetheless, these bots do have a key role in the social network \revision{to connect communities}, in particular, cross-cultural social marketing, bridging communal differences to better disseminate information \cite{moriuchi2021cross}.

\autoref{tab:networkmetrics} shows the network metrics measured for each type of user across the three locations of segregation. The metrics are extracted from the same all-communication network, and thus are comparable. In terms of \texttt{General Bots}, Chinese bots have the highest average eigenvector and total-degree centrality values. This indicates that the Chinese bots are the most well-connected within the network and have the highest degree of influence. \revision{This means that Chinese bots are better positioned to influence the network of users on the social media platform with their socio-political views and are thus more effective in disseminating their intended information, by means of their connectivity with a large number of users and influential users.} For \texttt{news bots}, Chinese bots have the highest eigenvector centrality while US bots have the highest total-degree centrality. Of note is that Chinese news bots have shown an average of 0 for betweenness centrality, or very negligible, which reflects that these bots do not aid in information flow within the network; they are likely existing mainly to push news stories. For \texttt{bridging bots}, Chinese bots have the highest eigenvector centrality, but bots from the reset of the world have the highest betweenness and total-degree centrality. This suggests that bridging bots from both US and China have rather little influence over the network.

\begin{table*}\centering
\begin{tabular}{|p{3cm}|p{2.6cm}|p{2.6cm}|p{2.6cm}|}
\hline
& \textbf{United States} & \textbf{China} & \textbf{Rest of the World} \\ \hline 
\multicolumn{4}{|l|}{\textbf{General Bots}}\\ \hline 
Eigenvector centrality & 4.32e-3 $\pm$ 9.47e-3 & 6.84e-3 $\pm$ 1.55e-2 & 3.38e-3 $\pm$ 8.05e-3 \\ \hline 
Betweenness centrality & 1.20e-7 $\pm$ 2.58e-6 & 2.31e-9 $\pm$ 2.03e-8 & 1.66e-9 $\pm$ 1.98e-8 \\ \hline 
Total degree centrality & 2.45e-6 $\pm$ 8.75e-6 & 2.62e-6 $\pm$ 5.25e-6 & 2.05e-6 $\pm$ 4.14e-6\\ \hline 
\multicolumn{4}{|l|}{\textbf{News Bots}}\\ \hline 
Eigenvector centrality & 1.04e-3 $\pm$ 2.78e-3 & 6.63e-3 $\pm$ 6.86e-3 & 8.02e-4 $\pm$ 2.63e-3 \\ \hline 
Betweenness centrality & 9.18e-9 $\pm$ 4.92e-8 & 0 $\pm$ 0 & 8.32e-9 $\pm$ 3.63e-8\\ \hline 
Total degree centrality & 4.27e-6 $\pm$ 1.24e-5 & 3.57e-6 $\pm$ 3.39e-6 & 5.47e-6 $\pm$ 1.20e-5\\ \hline 
\multicolumn{4}{|l|}{\textbf{Bridging Bots}}\\ \hline 
Eigenvector centrality & 5.99e-5 $\pm$ 4.11e-4 & 5.91e-4 $\pm$ 2.03e-3 & 2.54e-5 $\pm$ 6.55e-5 \\ \hline 
Betweenness centrality & 2.80e-9 $\pm$ 1.86e-8 & 6.95e-10 $\pm$ 3.55e-9 & 3.67e-9 $\pm$ 2.39e-8 \\ \hline 
Total degree centrality & 2.91e-6 $\pm$ 4.84e-6 & 1.98e-6 $\pm$ 1.86e-6 & 3.00e-6 $\pm$ 6.31e-6 \\ \hline 
\end{tabular}
\caption{Average Network Metrics of User Types}
\label{tab:networkmetrics}
\end{table*}

\begin{figure}[h!]
\includegraphics[width=\textwidth]{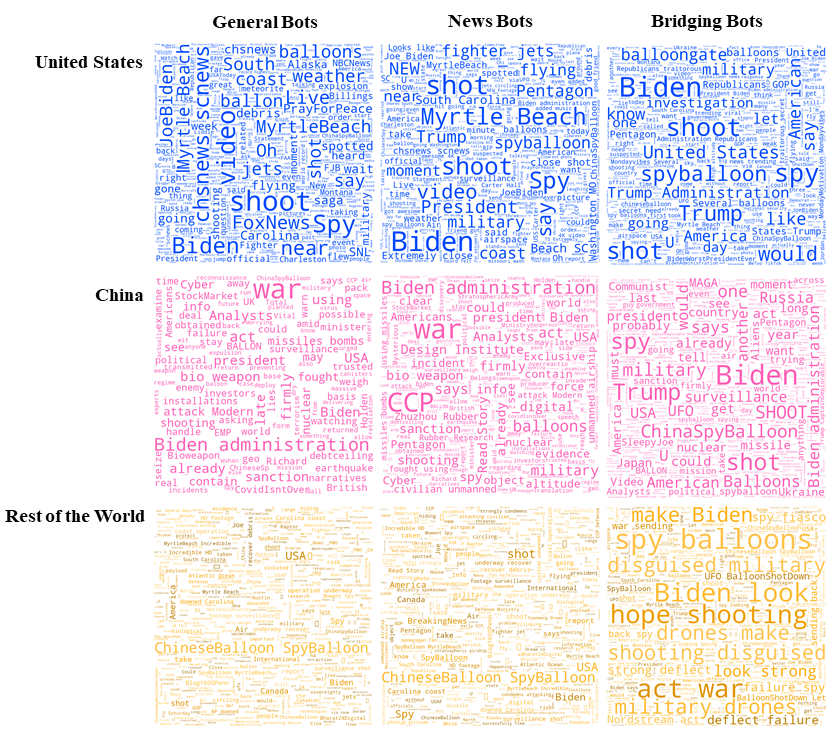}
\caption{Narrative themes reflected through topic analysis}
\label{fig:words}
\end{figure}

Next, we analyze the narrative themes reflects the narrative themes within the text of each type of user, and present the results in \autoref{fig:words}. Bots geotagged from the United States are generally concerned with the spatial location of the balloon and the possibility of it being a spy and surveillance balloon. Bridging bots from the United States discuss themes that directly relate political topics to the incident, for example discussion ``Trump Administration", ``Republicans", ``Biden". These terms reflect the perspective that US President Biden hesitated before taking action towards the balloon, and that there were similar balloons under the Trump Administration that were not discovered \cite{cnnChineseBalloons}.  One key phrase that arose is ``ballongate", a phrase closely linked to conspiracy narratives generally suffixed with -gate, like Pizzagate and Bridgegate \cite{tangherlini2020automated}. The term ``balloon-gate" has already gained an entry on Urban Dictionary as ``a political scandal, usually one with national security implications, in the making" \cite{balloongate}. This definition highlights the importance of this incident on political relations.

Bots geotagged from China discussed topics about war, sanctions, cyber warfare, investment and communism. These topics do not directly describe the balloon incident, but project possibilities of escalated hostility between both countries. It also reflects that Chinese bots view the actions of the US as a sign of aggression.

Bots from the rest of the world discuss a variety of topics, as seen from the generally similar-sized words throughout the word clouds, indicating that there are few extremely predominant topics within the discourse. These topics range from the responses from both countries, the shooting of the balloon, and also partaking in humorous theories such as the possibility of aliens and UFOs. 

\begin{figure}
\includegraphics[scale=0.65]{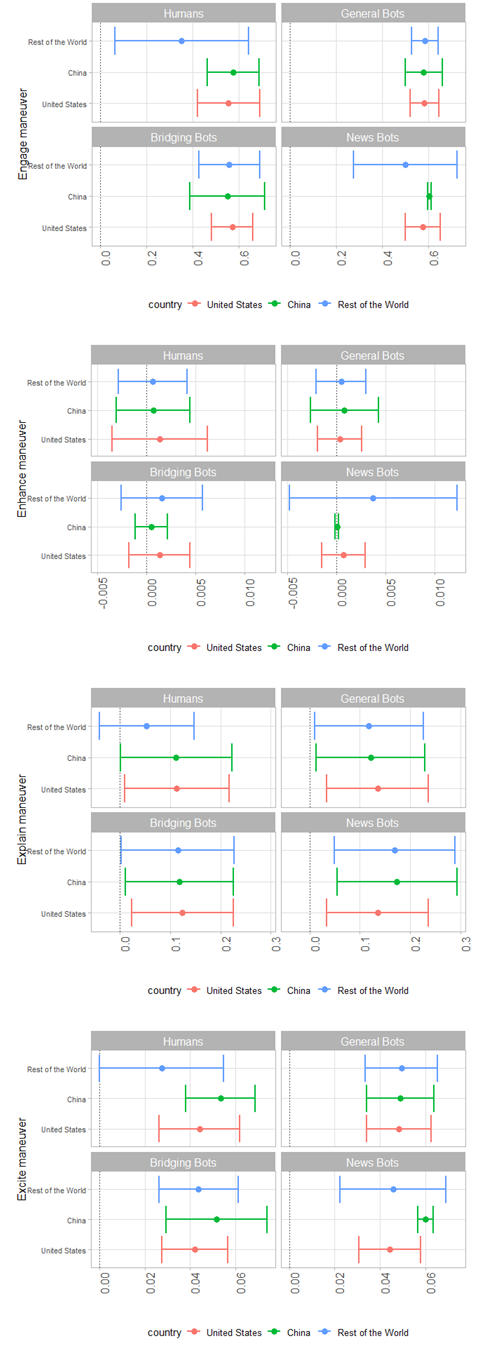}
\caption{Distribution of the E- Information Maneuver Metrics}
\label{fig:bend_e}
\end{figure}

\begin{figure}
\includegraphics[scale=0.65]{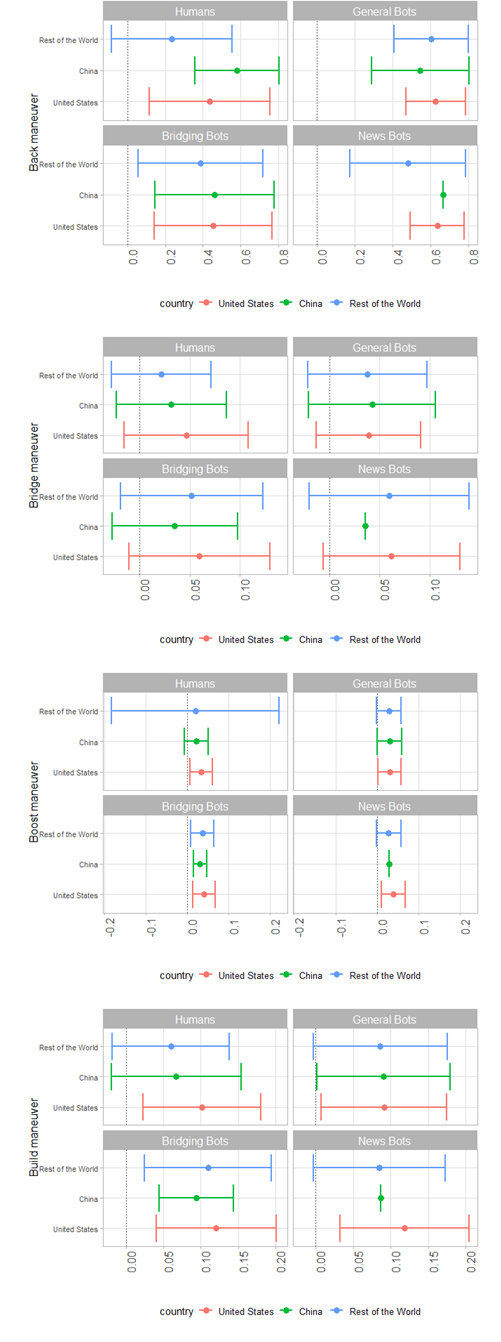}
\caption{Distribution of the B- Information Maneuver Metrics}
\label{fig:bend_b}
\end{figure}

\autoref{fig:bend_e} and \autoref{fig:bend_b} showcase the differences in the information maneuver tactics used by the different types of bots, measured using the BEND framework. In terms of the B- maneuvers (\autoref{fig:bend_b}), in which the users attempt to manipulate the social network, we find that overall, users perform the Back maneuver the most, followed by the Build, Bridge then Boost maneuver. This indicates that the bots are more concerned with supporting other users through likes and shares, building larger groups through @mentions and hashtags, rather than increasing the linkages between members. \texttt{General Bots} performed the most Back maneuvers, \texttt{Bridging Bots} performed the most Build maneuvers, \texttt{News Bots} performed the most Bridge maneuvers and \texttt{General Bots} performed the most Boost maneuvers.

In terms of the E- maneuvers (\autoref{fig:bend_e}), in which users attempt to manipulate narratives, we find that overall, users perform the Engage maneuver the most, followed by the Excite, then Enhance then Explain maneuver. This indicates that much of the narrative manipulation relies more on emotional appeal (Engage, Excite) rather than logical appeal (Explain). \texttt{General Bots} performed the most Engage and Explain maneuvers, \texttt{Bridging Bots} performed the most Enhance maneuver and \texttt{News Bot} performed the most Excite maneuver.

\section{Discussion}
In this political discussion, we observe that that average proportion of bots is 47.30\%. This is higher than the average observed in past studies, which ranged from 5 to 18\% \cite{tan2023botpercent,fukuda2022estimating}. This could be due to a few reasons: the bot percentage in political communities are generally much higher than other communities like entertainment \cite{tan2023botpercent}, or that this is a diplomatic event that involves two major powers thus gathers more interest, and hence groups of actors are compelled to deploy more bots in an attempt to control the narratives, or that the event is extremely juicy, which can be evidenced by the large number of memes, jokes and taglines that have been constructed online \cite{kutllovci2023s}.
The differences in the proportion of bots that are geotagged to be from across US, China and the rest of the world indicates that the distribution of Twitter bots are not spatially homogeneous. This provides leads towards the countries that more bots originate from, which could be possible signs of state-sponsored bots, which is especially important in managing the country's image in a political narrative, keeping in line with the intent of diplomacy.

One key issue about the distribution of users to note is that while we have segregated users based on their geolocation (US, China, Rest of the World), these users \revision{are} not necessarily physically originating from the identified geolocated country. Instead, these reflect their association with the country, and by extension our analyses reflect the views of users that affiliated themselves with a particular country. This is important to note because Twitter is banned in China due to the Chinese firewall, yet there are users that reflect a geolocation of China. These users therefore, must be users that are actually geolocated in China, or users that affiliate their geolocation to China. Either way, we unfortunately are unable to differentiate the truthfulness of the geolocation of Twitter users through the information in the data collected, and therefore rely on the self-presentation of location. \revision{Our analysis of only English language tweets also has value to the users that are geo-tagged to China. It reflects the thoughts that users affiliated with China are projecting towards the English-speaking community, showcasing the projection of diplomatic narratives and the projection of soft power.}

Across the three main cultural groups (United States, China, Rest of the World), we observe that the same type of social bots exist and play the same roles. The General Bot is the most common bot type that is present throughout geotagged bot accounts from all countries, followed by News Bots. The presence and employment of Bridging Bots are generally rare throughout all cultures, possibly because it takes additional effort to program the automated accounts to identify multiple groups of people and link them together. 

Within this political event, we identified an average of 3\% of the bot users are \texttt{News Bots}. This is consistent with previous studies where bots are observed to be used as active news promoters during critical events, such as the 2020 Coronavirus pandemic \cite{al2020bots}. The average sentiment of \texttt{news bots} is neutral, indicating that the bots do not inject much emotion into their posts, and thus do not actively attempt to manipulate opinions. In contrast to the prevailing conception that social media bots are generally malicious bots (80\% of Americans who have heard about social media bots view them as malicious \cite{pewresearchMostAmericans}), this analysis of \texttt{news bots} show that there are bots that are helpful bots: in this case, \texttt{news bots} serve to disseminate news.

The proportion of \texttt{bridging bots} are extremely rare, with \texttt{bridging bots} that originate from all three regions being less than 1\%. Despite their small presence within the social media space, these bots play an important role. They enable information flow across boundaries, be it geographical, language or ideological, allowing users from one community to receive information and opinions from the other community. This pattern has also been observed on Twitter during the Arab Spring revolution in Egypt in 2011, where \texttt{bridging bots} enabled information flow between Arabic and English language spheres \cite{bruns2013arab}.  In our study, bridging bots enabled information flow between China and US bots, promoting the exchange of topics and emotions.

However, the employment of these bots across the cultures differ. Bots from different countries are used to present different narrative themes and frames. This most often aligns with the country's diplomatic perspective of the event. For example, News Bots from the United States focused on the surveillance aspect of the balloon, while those from China focused on the comments from the Biden administration.

Bots within this diplomatic incident mainly use positive information maneuvers, in line with image-enhancing projection techniques in diplomacy \cite{sterling2018new}. To expand the reach of their message through the network, bots mostly use the Back and Build maneuvers. These maneuvers make use of actions like @mentions, likes and retweets, which can easily be automated through the Twitter API. 
These observations are similar to previous studies where bots are observed to heavily use the positive B- maneuvers to support their stance towards/against the Palestine-Israel crisis \cite{danaditya2022curious}. To increase readership of their message, \texttt{General Bots} and \texttt{Bridging Bots} make use of emotional appeal using the Engage and Excite maneuvers, rather than logical appeal. This technique has been observed in past studies where bots target emotionally appealing topics and are extremely effective with using emotions as part of their persuasive dialogue in both political topics \cite{nonnecke2022harass} and general discourse \cite{paavola2016understanding}. \texttt{News Bots} attempt to make their short news headlines catchy, as evidenced by their high Excite maneuver. 



While we have segregated users by geography, we did not investigate evidence of state manipulation, and therefore are unable to comment on whether the users are state-manipulated actors. There is, however, a possibility that a good amount of bot accounts are state-sponsored accounts, be it originating from the US, China or the rest of the world, with intentions to put forth certain narratives that are in line with the diplomatic concerns of the state. 

\revision{Further, we} do not suggest that our typology of bots in this article is exhaustive. There are other types of bots that we have not explored within this discourse that exists in social media. Such bots include amplifier bots which have been shown to intensify opinions in political discussions \cite{mckelvey2017computational,yoon2022super}. Another is spam bots which have been observed to spread phishing links on Twitter \cite{chu2010tweeting} and overwhelm Twitter users with messages spreading their own ideology \cite{jamison2019malicious}.

\revision{Our study has broader implications towards policymakers as well. Based on the geolocated discourse, we can understand the narratives for projection of political images used by automated agents affiliated with region. The interactivity on digital platforms provides deeper understanding into the portrayal of a country on social media, and the perception of political issues by discussants on online social media. With these information, those in policy and governance can systematically identify public sentiment, which can be useful in devising communication strategies to address these sentiments. Those involved in studying foreign affairs can better identify a country's projected national image and stance towards the political event, and analyze and anticipate possible political responses.}

\subsection{Limitations and Future Work}
We highlight a few limitations of our work. Naturally, our discussion is not exhaustive and a few limitations nuance the discussions of the work.

One key limitation of this study is the data source: data collected from the Twitter API only presents 1\% of the discourse in the social media space, and thus we make the assumption that our collected data represents the social media discussions. Our study is limited to the discussion on Twitter which is less popular in China because it is a restricted platform \cite{sullivan2012tale}. Users with more extreme opinions are typically more vocal on social media, thus we suggest caution in extrapolating the findings, and be cautious of the silent majority/ vocal minority effect \cite{ng2022pro}. For these reasons, the study does not reveal the full scope of the inauthentic bot activity stemming from both countries during this event. Research of wider breadth is needed to map out the bot campaign and reach of both US and Chinese bot accounts.

Additionally, geolocation identification of user accounts relies on self-presentation of the accounts, and Twitter users can erronously reflect their location. Some do so for humor, like specifying itself from China and having a profile picture of a balloon along with a description ``just floating around in cyberspace before being shot down"; others do so to deliberately avoid association with specific countries; and yet others do not provide any location information.

\revision{Our bot detection algorithm relies on a supervised machine learning algorithm that is trained on a manually labeled dataset. It extracts bot-like features such as extremely frequent retweeting behavior and temporal periodicity of posting. These general-purpose bot detection algorithms have been found to be prone to error \cite{hays2023simplistic}, which can therefore affect the observations. In particular, some users identified are the highly engaged user, the user that tweets and retweets extremely often such that the behavior comes across as bot-like to the algorithm. Also, in our determination of whether the user is a bot or not, we used the 0.70 threshold value, based on past systematic large-scale analyses on threshold values \cite{ng2022stabilizing}. However, further work should be done to measure the impact of different thresholds on the results and on the efficiency of extracting different types of bots. }

And lastly, our analysis includes only English language tweets. While this showcases what the English speaking community in both the US and China are discussing about, this precludes the opinions of the Chinese speaking community, a community that is native to the country China. 

Following from the limitations, we suggest directions for future work. 
To cope with the geolocation issue, some researchers have attempted to infer a user's geolocation from the places mentioned within tweets or the style and language of the tweets \cite{han2014text}. Other methods make use of the social network of the user, inferring the user's geolocation via their friends \cite{jurgens2015geolocation}. These methods can be employed to enhance the identification of a user's geolocation via their declared location, and result in a larger number of users that are available for analysis.

We also suggest in-depth investigations of other types of bots that are used for digital diplomacy on social media, expanding the scope of analysis of automated agents that are employed in the realm of cross-country diplomacy. An example of a type of bot that warrants investigation is the propaganda bots, which have been observed during the Gulf crisis to spread propaganda and fake news \cite{jones2019gulf}.

Downstream work looks towards performing cross-lingual analysis within a diplomatic event that spans more than one country. This is especially important in understanding the bot activity in social media communities other than English. Particularly, in this event it relates to the Chinese speaking community on Twitter: Chinese language bot activity has increased over the years, and bot accounts are observed to be bombarding searches for Chinese cities with tweets related to pornography and gambling \cite{techcrunchDespiteBan}. \revision{Analysis of bot-like users within a diplomatic event also includes in-depth investigation into the background intent of these users: sets of inauthentic accounts created and operated by political actors working to influence social discourse, sets of programmed inauthentic accounts that aggresively amplify and disseminate information and the sets of highly engaged and active users that possibly monetize their content or linked content from tweet. Finally, further work can be done to analyze the temporal changes in the usage of information maneuvers, extending our work from a wrinkle in time towards a cinematic replay of the variants of maneuver strategies deployed through time.}

\section{Conclusion}
The topic of automation and online politics have become a major area of investigation in computational social science because the spread of inauthentic bot accounts on social media can pose a problem. 

This article is the first to develop methodologies to pick out certain types of bots that are used during a diplomatic incident. These methodologies are not restricted to the incident in the study, nor are they restricted to diplomatic events; they are generic methodologies that can be applied to identifying bots across a variety of events in the social media space. Following which, this article provides the first academic insights into the nature of three types of bots during a diplomatic incident, identifying the differences in the topics and interactions between each type of bot in relation to the country it likely originates from.

We find that all three types of bots examined within this article -- General Bots, News Bots and Bridging Bots -- are present in our regions of study, namely the United States and China. With regards to the balloon incident which is a diplomatic situation between the two world powers, the bots tweet on different topics. US bots are more interested in the location of the balloon and the eventual shooting down of the balloon, while Chinese bots view the actions by the US as aggression.

All three types of bots engage in the positive narrative and network information maneuvers, but bots from different countries perform different degrees of each maneuver. The most performed narrative maneuver is the Back maneuver as bots prop each other's messages up, while the most performed narrative maneuver is the Engage maneuver as bots actively engage each other through network communication interactions. 

As bots become more prevalent in the digital space, they can be used to communicate with the general public (e.g., \texttt{news bots}). They should also be monitored to understand the general public, as \texttt{general bots} and \texttt{bridging bots} can potentially alter the public opinion. Overall, this article analyzed the political participation by automated users. Through analysis of the narratives and the network influence of social media bots affiliated with two major powers, we contribute to the literature on soft power and online diplomacy, and present an understanding of the image that users affiliated with each country are projecting on social media. We hope that the article will set forth discussions surrounding the use of bots in digital diplomacy.

\section{Ethical Statement}
There are several ethical points to consider in our work.

In this study, we only extracted publicly available data using the Twitter API, and no attempt was made to retrieve protected tweets. Within this article, we do not process the usernames or unique personal identifiers of the social media accounts during our analysis. During our analysis, we use only the aggregate trends and do not investigate the profile and activity of individual users. We do not mention specific user names within our report because many of these accounts are still active online.

The conclusions that we obtained through applying our methodology are based of the observations of the Twitter accounts and their posts. This represents but a slice of the discussion online and do not necessarily represent the entire population. Therefore, one must be cautious in using our insights to inform behavior or policy.


\begin{backmatter}

\section*{Abbreviations}
JSON: JavaScript Object Notation \\
US: United States \\ 
UFO: Unidentified Flying Object \\
URL: Unified Resource Locator

\section*{Data Availability}
The original set of data collected for this article can be obtained from \url{https://drive.google.com/drive/folders/1RDH7l2CYsr-DbnW1Si1LGZPX_KcYUE9Z?usp=drive_link}.
Further data used in this article can be obtained from the corresponding author in accordance to Twitter's Terms and Conditions. 

\section*{Competing interests}
The authors declare that they have no competing interests.

\section*{Funding}
This material is based upon work supported by the U.S. Army Research Office and the U.S. Army Futures Command under Contract No. W911NF-20-D-0002, Office of Naval Research (Bothunter, N000141812108), Scalable Technologies for Social Cybersecurity/ARMY (W911NF20D0002), Air Force Research Laboratory/CyberFit (FA86502126244). The content of the information does not necessarily reflect the position or the policy of the government and no official endorsement should be inferred.

\section*{Author's contributions}
LHXN: Conceptualization, methodology, analysis, writing. KMC: Review and editing. All authors have read and agreed to the published version of the manuscript.

\section*{Acknowledgements}
We thank our colleagues at the IDeaS lab at Carnegie Mellon University for their initial feedback on the ideas of this work. We also sincerely thank the reviewers for providing great comments that made this piece a much better one.


\bibliographystyle{bmc-mathphys}
\bibliography{references} 

\section{Figure/ Table Legends}
Figure 1: Overview of Methodology for Identifying Types of Bots

Figure 2: Methodology for identification of News Bots

Figure 3: Methodology for identification of Bridging Bots

Figure 4: Proportion of User Types 

Table 1: Examples of Tweets by Bot Types. User names and URLs are removed as these can potentially uniquely identify the users

Figure 5: User Type Distribution through Social Network Analysis 

Table 2: Average Network Metrics of User Types 

Figure 6: Narrative themes reflected through topic analysis 

Figure 7: Distribution of E- Information Maneuver Metrics

Figure 8: Distribution of the B- Information Maneuver Metrics

\end{backmatter}
\end{document}